\documentclass[fleqn,usenatbib]{mnras}

\usepackage[T1]{fontenc}
\usepackage{ae,aecompl}

\usepackage{graphicx}	
\usepackage{amsmath}	
\usepackage{amssymb}	

\title{Constraining Effective Temperature, Mass and Radius of Hot White Dwarfs}

\author[E. do A. Soares]{
Elvis do A. Soares\thanks{E-mail: esoares@if.ufrj.br}
\\
Instituto de F\'isica, Universidade Federal do Rio de Janeiro, C.P. 68528, Rio de Janeiro 21945-970, RJ, Brazil
}

\date{Accepted XXX. Received YYY; in original form ZZZ}

\pubyear{2017}

\begin{document}
\label{firstpage}
\pagerange{\pageref{firstpage}--\pageref{lastpage}}
\maketitle

\begin{abstract}
By introducing a simplified transport model of outer layers of white dwarfs we derive an analytical semi-empirical relation which constrains effective temperature-mass-radius for white dwarfs. This relation is used to classify recent data of white dwarfs according to their time evolution in non-accretion process of cooling. This formula permit us to study the population map of white dwarfs in the central temperature and mass plane, and discuss the relation with the ignition temperature for C-O material. Our effective temperature-mass-radius relation provide a quick method to estimate the mass of newly observed white dwarfs from their spectral measurements of effective temperature and superficial gravity.
\end{abstract}

\begin{keywords}
stars: evolution -- stars: atmospheres -- white dwarfs -- radiative transfer
\end{keywords}



\section{Introduction}

White dwarfs (WD) are frequently treated in the zero temperature approximation of highly degenerate electron gas pressure, which provide an adequate description of the stellar structure as a whole. On the other hand, in the area close to the surface we know that the observed white dwarfs have considerable high effective temperatures ($T_\text{eff}$) ranging from 5,000 K to over 100,000 K, losing their thermal energy emitting radiation  \citep[e.g.,][]{Koester1990}. The external layers, although the amount of mass contained there is small, determine the thermal evolution of the whole white dwarfs, while the bulk degenerate electrons keep the star's core essentially isothermal due to its high thermal conductivity and hotter than the crust. The radiative opacity in the outermost layers prevents the white dwarf to cool quickly. Comparison between observations and models requires the corrections from the finite temperature effects to the white dwarf stellar structure.

Many works have been developed, since the discovery of the maximum mass of ideal white dwarfs (WD) by \cite{Chandrasekhar1931TheMass}, in the field of finite temperature corrections to the degenerate equation of state (EoS) \cite[e.g.,][]{Marshak1940TheStars.,Hubbard1970HotDwarfs,DeCarvalho2014}. However, as commented in \cite{Boshkayev2016EquilibriumTemperatures}, a systematic analysis using empirical mass-radius relations obtained from the spectroscopic or photometric measurements of masses and radii is still needed to understand the precise structure and the dynamics of time evolution of WDs.

Moreover, the total number of observed hot ($T_\text{eff} >$ 10 000 K) white dwarfs (WDs) has increased enormously mainly due to the Sloan Digital Sky Survey (SDSS; \citealt{Eisenstein2006A4}). Follow-up high quality ground-based spectroscopy of survey objects yield large samples of hot WDs with precise measurements of effective temperatures and superficial gravities. More than 10,000 spectroscopically identified white dwarfs with determined effective temperatures ($T_\text{eff}$) and superficial gravities ($\log g$) have been detected to date \citep{Kepler2007,Kleinman2013}, giving us the opportunity to explore the white dwarf mass distribution, which ultimately provides insights into mass-loss processes during stellar evolution and the mass budget of the Galaxy.

Beyond that, from the astrophysical point of view, hot WDs are important: (i) to elucidate the evolutionary links between WDs and their pre-white dwarf progenitors, i.e., whether they are from the asymptotic giant branch (AGB), the extended horizontal branch, stellar mergers, or binary evolution. (ii) to understand their roles in the process of chemical evolution of the Galaxy, because white dwarf progenitors lose their outer layers which are carbon, nitrogen, and oxygen rich at the top of the asymptotic giant branch (AGB) (iii) to improve our knowledge of type Ia supernova events, with important underlying implications for cosmology \citep[e.g.,][]{Hillebrandt2013}.

The main goal of this work is to use the data from Sloan Digital Sky Survey (SDSS) Data Release 7 to constrain the effective temperature, mass and radius of hot white dwarfs in a simple analytical relation, introducing a very simplified thermal transport model of the outer layers. The present paper is organized as follows. First we introduce the extended model for outer radiative layers in \autoref{sec: Extended Model for Outer Radiative Layers}. Based on this model, we derive an analytic relation among the effective temperature, mass and radius containing two parameters, which related to the the transport properties of the outer layers. In \autoref{sec: Semi-Empirical relation among Mass, Radius and Effective Temperature} by using the \emph{SDSS}-DR7 data, we determine these two parameters as function of WD's mass. In \autoref{sec: Semi-Empirical Mass-Radius Relation} using thus obtained semi-empirical mass-radius relation for WD for each effective temperatures, we discuss the implications of these constraints for central temperature and ignition of the nuclear material inside WD in \autoref{sec: Central Temperature and Nuclear Ignition}. In \autoref{sec: Estimating Masses} we use our results to estimate masses and radii for others observed hot white dwarfs, as in the \emph{Gaia} DR1. In \autoref{sec: Conclusions}, we discuss our results from physical point of view and perspectives for the further study.

\section{Extended Model for Outer Radiative Layers of White Dwarfs}
\label{sec: Extended Model for Outer Radiative Layers}

The highly degenerated electron gas inside a WD provides a high thermal conductivity as a result of the large mean free path of the degenerate electrons in the filled Fermi sea \citep[e.g.,][]{Cox2004}. Such high thermal conductivity together with the lack of nuclear reactions do not allow large temperature gradients, leading to an almost uniform temperature in the WD interior. On the other hand, in the domain close to its surface, the density $\rho$ decreases and the matter becomes quickly non-degenerate. Then, the dominant heat transfer is the radiative one (and a little convection), and the heat conduction becomes much smaller if compared to the degenerate electron gas. Therefore, we expect that the structure of a WD can be modeled as an isothermal core covered by non-degenerate surface layers which isolates the degenerate core from the outer space \citep[e.g.,][]{Kippenhahn2012}.

To exploit the above image, let us introduce a simple approach to describe the energy transfer mechanism in the outer layers of a white dwarf. The main simplification consists in attributing the outermost layers as the region responsible by the thermal regulation of the white dwarf and the core responsible for the mechanical regulation of the stellar structure, i.e., the core is in a hydrostatic equilibrium and the outer layers is in a stationary state of radiative energy transfer. The outer layer region starts where the degenerate matter becomes non-degenerate (ideal gas) matter. The thermal gradient and the hydrostatic equilibrium are maintained by 
\begin{equation}
\frac{dT}{dr} = - \frac{3}{4ac} \frac{\kappa \rho}{T^3} \frac{L_r}{4\pi r^2}
\label{equacao do calor}
\end{equation}
and
\begin{equation}
\frac{dP}{dr} = - \frac{G M_r}{r^2} \rho,
\end{equation}
where $M_r = \int_0^r \rho 4 \pi r^2 dr$ and $L_r = \int_0^r \epsilon 4 \pi r^2 dr$. Dividing one equation by the other, we can write
\begin{equation}
\frac{dP}{dT} = \frac{16\pi ac G}{3}\frac{1}{\kappa} \left(\frac{M_r}{M}\right) \left(\frac{L}{L_r}\right) \frac{M}{L}T^3
\label{pressao temperatura gradiente}
\end{equation}
where $M$ and $L$ are respectively the mass and luminosity of the star. This equation can be integrated with the hypothesis that the outer layers are too thin to contribute to the mass, i.e., $M_r\approx M$ and there is no energy generation ($\epsilon =0$) in these layers, i.e., $L_r\approx L$. By supposition, the material here is a non-degenerate and fully ionized gas, so we can use the ideal gas EoS and the Kramers opacity, $\kappa=\kappa_0 \rho T^{-3.5}$. With this we have
\begin{equation}
\rho^{2} = \left(\frac{2}{8.5} \frac{4ac}{3} \frac{4\pi G M}{\kappa_0 L} \frac{\mu}{N_A k_B}\right) T^{6.5},
\label{relacao Shapiro}
\end{equation}
integrating the Eq.\eqref{pressao temperatura gradiente} from $P=0$ when $T=0$. The Eq.\eqref{relacao Shapiro} is a well-known result, as can be seen in \cite{Shapiro1983}.

Using the Eq.\eqref{relacao Shapiro}, which is valid for any $r$ inside the outer layer, we can eliminate $\rho$ from Eq.\eqref{equacao do calor} and integrate this equation from an effective radius (region where the photons decouple from surface) where the temperature is the effective temperature to the external radius of the WD where the temperature is zero, we get 
\begin{equation}
T_\text{eff}= \left(\frac{1}{4.25}\frac{\mu}{N_A k_B}  \right) \frac{G M}{R} \left( \frac{R}{R_\text{eff}} -1\right).
\label{Shapiro inspired}
\end{equation}

It is reasonable to assume that the effective radius can be related with the Chandrasekhar radius, $R_\text{eff}=\xi R_\text{ch}(M)$, with $\xi \sim 1$, since the effective radius should be very close to the core surface. Then
\begin{equation}
T_\text{eff} = (588,\!862 \text{ K}) \mu \left(\frac{M}{M_\odot} \right) \left(\frac{R}{R_\oplus} \right)^{-1} \left[ \left(\frac{R}{\xi R_\text{ch}(M)}\right)-1 \right],
\label{Semi-empirical radius-effective temperature relation for hot white dwarfs}
\end{equation}
where $\mu$ and $\xi$ are parameters to be determined. For simplicity we further use the analytical approximated expression for the radius of ideal white dwarfs $R_\text{ch}(M)$ given by \cite{Nauenberg1972AnalyticStars} as
\begin{equation}
R_\text{ch}(M) = \frac{2.45354}{\mu_e} R_\oplus \left(\frac{M}{M_\text{ch}}\right)^{-1/3} \left[1-\left(\frac{M}{M_\text{ch}}\right)^{4/3} \right]^{1/2}
\end{equation}
with $M_\text{ch} = 5.816\; M_\odot /\mu_e^2$ and $\mu_e$ is the mean molecular weight per electron. 

The Eq.\eqref{Semi-empirical radius-effective temperature relation for hot white dwarfs} determines the effective temperatures of the white dwarf stars as a function of their masses and their radii. This relation is semi-empirical because the correction parameters $\mu$ and $\xi$ are fitted by data, but it is physically based on the model of transport phenomena in the outer layers of the white dwarf stars.

\section{Semi-Empirical relation among Mass, Radius and Effective Temperature}
\label{sec: Semi-Empirical relation among Mass, Radius and Effective Temperature}

The hot WD are composed mainly by carbon (C) and oxygen (O), but their observed spectra show us that their atmosphere is dominated by hydrogen (H) or helium (He), with the dominant element almost thousand times more abundant than the other elements in that location. For these reasons, we expect then $\mu_e=2$ for the core of both DA-WD and DB-WD due to their internal composition, but the parameters $\mu$ and $\xi$ of DA-WD and DB-WD must be different.

\subsection{Data Set and Procedure}

Using the data from the Sloan Digital Sky Survey (SDSS) Data Release 7, \cite{Kleinman2013} reported 42,154 spectroscopically confirmed white dwarf stars. From the 14,120 clean DAs classified by the spectra, we use the 2,216 stars with $S/N \geq 15$ and $T_\text{eff} \geq $ 13,000 K. Of the 923 stars which they classified as clean DBs, we use the 140 stars with $S/N \geq 15$ and $T_\text{eff} \geq $16,000 K. 

The masses of the identified clean DA and DB stars are calculated from the effective temperature $T_\text{eff}$ and superficial gravity $g$ values obtained by spectra. These relations are based on full evolutionary calculation of hydrogen-rich DA white dwarfs and hydrogen-deficient DB white dwarfs, as discussed in that paper. These evolutionary sequences constitute a complete and homogeneous grid of white dwarf models that captures the core features of progenitor evolution, in particular the internal chemical structures expected in the different types of white dwarf stars.

Therefore we will not consider the data uncertainties to fit the parameters $\mu$ and $\xi$, assuming that such cannot be divided in systematical uncertainties from the stellar evolution simulations, and observational uncertainties coming from the spectra. Of course, this procedure must be revised in the future when the systematical uncertainties were recognized.

Then, to adjust the parameters $\mu$ and $\xi$ we choose a narrow range of masses around a mean value $M$ with a width of $0.001 M_\odot$ sufficient to determine the parameters for a single value of mass, because white dwarf stars with same mass must evolve similarly. 

\subsection{DA white dwarfs}
\label{DA white dwarfs}

\begin{figure}
\centering
  \includegraphics{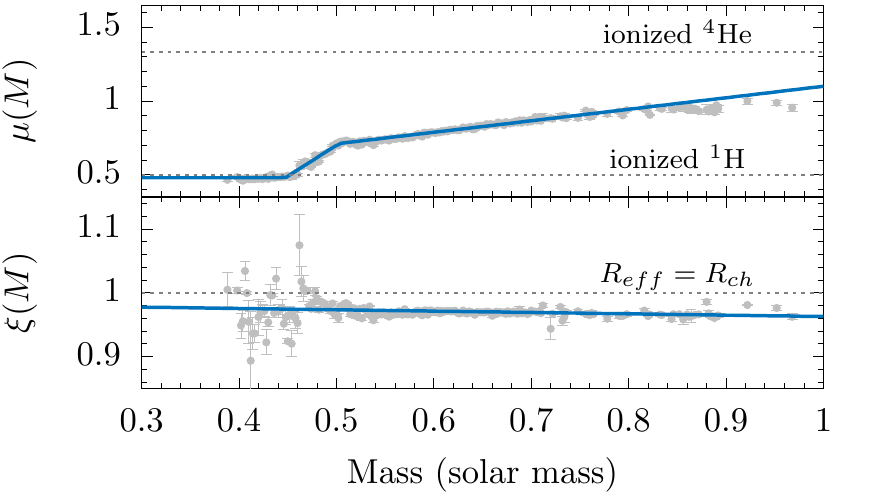}
  \caption{Parameters $\mu$ and $\xi$ as a function of mass for DA-WD. The top panel shows two references for $\mu$ as dotted lines, a pure ionized He gas and a pure ionized H gas. In the bottom panel we represent the case where $R_\text{eff}=R_\text{ch}$ as dotted line.}
  \label{fig: Parametros mu e xi - DA}
\end{figure}

For the hydrogen-rich DA white dwarfs, the main behavior of the coefficients $\mu(M)$ and $\xi(M)$ can be fitted by the simplest forms
\begin{equation}
\mu(M) = 
\begin{cases} 
      0.48(0.02) & M/M_\odot < 0.448 \\
      4.2(0.2) \frac{M}{M_\odot}-1.4(0.1) & 0.448\leq M/M_\odot \leq 0.503 \\
      0.78(0.01) \frac{M}{M_\odot}+0.32(0.01) &  M/M_\odot > 0.503
   \end{cases}
 \label{eq: parameter mu DA-WD}
\end{equation}
and
\begin{equation}
\xi(M) = 0.984(0.002)-0.021(0.003)\frac{M}{M_\odot},
\label{eq: parameter xi DA-WD}
\end{equation}
according with the Figure \ref{fig: Parametros mu e xi - DA}, and the uncertainties are represented in the parentheses.

A transition from a pure hydrogen composition for a hydrogen-helium mixture in the outer layers is presented by the parameter $\mu(M)$ in the top panel of Fig.\ref{fig: Parametros mu e xi - DA}. This outer layer composition transition is marked by an core composition transition. In fact, WD with mass below 0.452 $M_\odot$ are helium-core white dwarf stars \citep{Althaus2009EvolutionProgenitors} and WD with mass above 0.452 $M_\odot$ are carbon-oxygen white dwarf stars \citep{Althaus2005Mass-radiusStars}. 

\begin{figure}
\centering
  \includegraphics{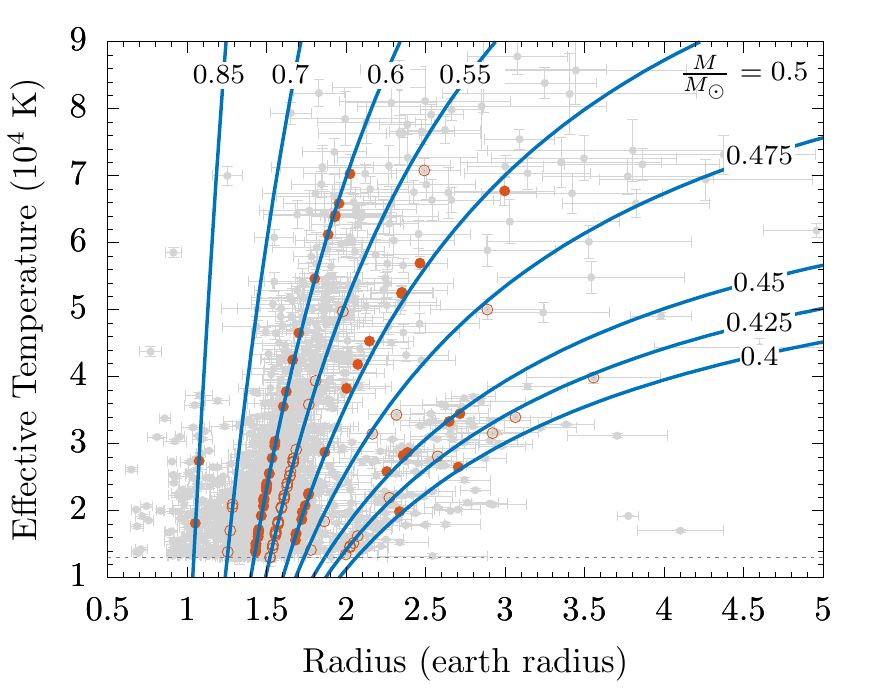}
  \caption{Semi-empirical radius-effective temperature relation for DA-WD with different masses. The orange circles represent some values of mass and the blue lines are their correspondent fits. The light gray circles are the available data for DA-WD from the SDSS-DR7.}
  \label{fig: Relacao massa temperatura DA}
\end{figure}

The Figure \ref{fig: Relacao massa temperatura DA} illustrates the radius-effective temperature of DA-WD with different masses using the Eq.\eqref{Semi-empirical radius-effective temperature relation for hot white dwarfs} and the parameters $\mu(M)$ and $\xi(M)$ given by Eqs.\eqref{eq: parameter mu DA-WD} and \eqref{eq: parameter xi DA-WD} for several values of WD masses (blue lines). The orange circles represent the correspondent data for these mass values. The light gray circles are the available data for DA-WD in the SDSS-DR7. Our analytic lines are not displayed for all data but there is a excellent agreement between the semi-empirical relation and the DA-WD data for masses above $0.4\; M_\odot$. Bellow this mass value, the parameters cannot be adjusted because the low statistics of the data.

\subsection{DB white dwarfs}
\label{DB white dwarfs}

\begin{figure}
\centering
  \includegraphics{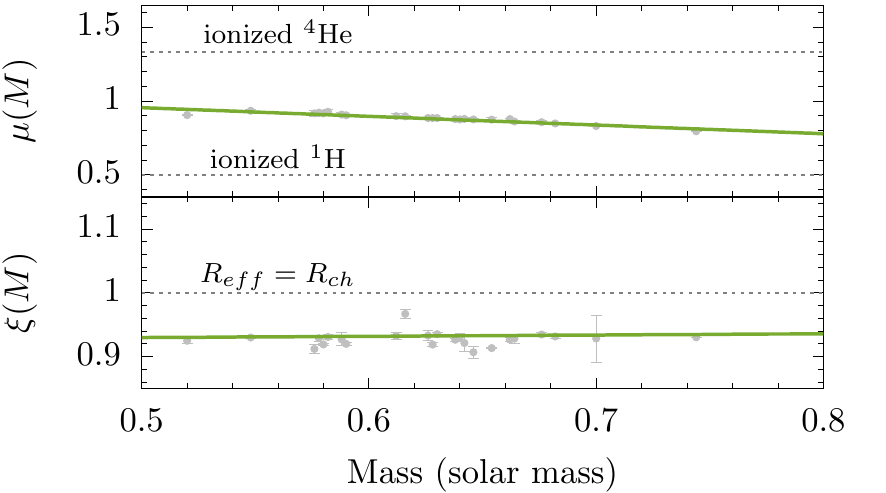}
  \caption{Parameters $\mu$ and $\xi$ as a function of mass for DB-WD. The top panel shows two references for $\mu$ as dotted lines, a pure ionized He gas and a pure ionized H gas. In the bottom panel we represent the case where $R_\text{eff}=R_\text{ch}$ as dotted line.}
  \label{fig: Parametros mu e xi - DB}
\end{figure}

For the hydrogen-deficient DB white dwarfs, the main behavior of the coefficients $\mu(M)$ and $\xi(M)$ can be fitted by the simplest forms
\begin{equation}
\mu(M)= 1.25(0.03) - 0.59(0.05) \tfrac{M}{M_\odot},
\end{equation}
\begin{equation}
\xi(M)= 0.92(0.03)+0.02(0.05) \tfrac{M}{M_\odot},
\end{equation}
according with the Figure \ref{fig: Parametros mu e xi - DB}.

For hydrogen-deficient white dwarf stars with stellar mass values from 0.515 to 0.870 $M_\odot$ \citep{Althaus2009NEWEVOLUTION} there is not core composition transition, and consequently the parameter $\mu(M)$ varies slowly in this range of mass values. 

\begin{figure}
\centering
  \includegraphics{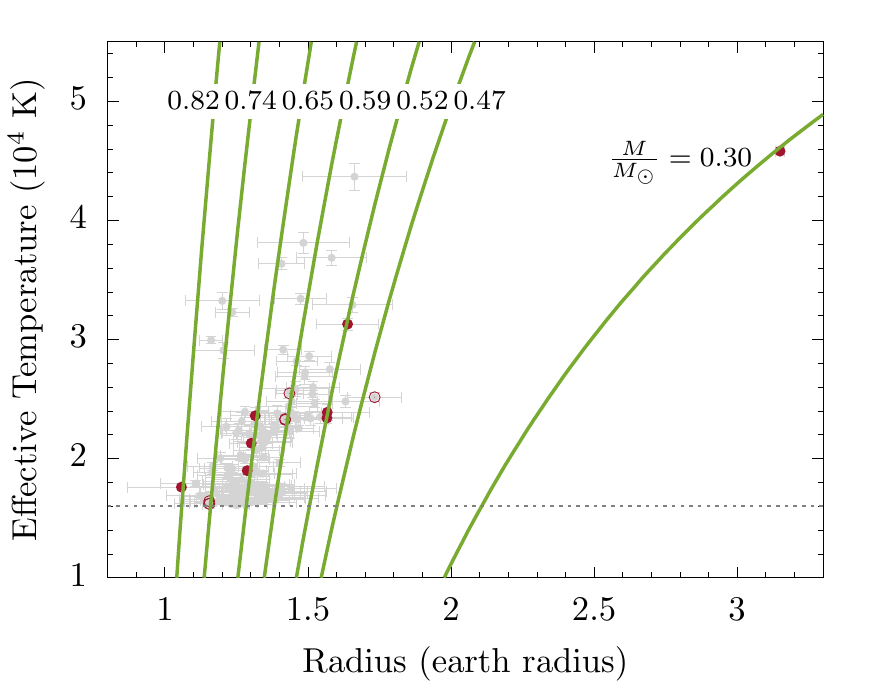}
  \caption{Semi-empirical radius-effective temperature relation for DB-WD with different masses. The red circles represent some values of mass and the green lines are their correspondent fits. The light gray circles are the available data for DB-WD from the SDSS-DR7.}
  \label{fig: Relacao massa temperatura DB}
\end{figure}

Alike the case of DA-WD, the Figure \ref{fig: Relacao massa temperatura DB} illustrates the radius-effective temperature of DB-WD with different masses using the Eq.\eqref{Semi-empirical radius-effective temperature relation for hot white dwarfs} and the parameters $\mu(M)$ and $\xi(M)$ given in this section. The red circles represent some mass values for which the semi-empirical relation is calculated, represented by the green lines. The light gray circles are the available data for DB-WD in the SDSS-DR7. There is again a good agreement between the semi-empirical relation and the DB-WD data despite the low statistics. Note that the very isolated point for $M=0.3 M_\odot$ is stay also well on our curve.

\section{Semi-Empirical Mass-Radius Relation}
\label{sec: Semi-Empirical Mass-Radius Relation}

The mass-radius relation is a fundamental ingredient to understand the physics of white dwarfs. The first mass-radius relation given by \cite{Hamada1961ModelsStars.} assumed a zero temperature fully degenerate core. Finite temperature corrections to C and O nuclear material and the non-degenerate outer layers of He and H were included by \cite{Althaus2010EvolutionaryStars}. Recently, \cite{Holberg2012OBSERVATIONALRELATION} constraint the degenerate mass-radius relation with the observations, but there is a doubt about the favored models to estimate the mass and radius of WD, using "thick" H envelopes or "thin" H envelopes. Therefore, the mass-radius relation of white dwarfs is not greatly constrained by observations.

Our constraining relation Eq.\eqref{Semi-empirical radius-effective temperature relation for hot white dwarfs} can be inverted to give
\begin{equation}
R = \xi R_\text{ch}(M) \left[ 1- \frac{1}{\mu} \left(\frac{T_\text{eff}}{T_0}\right) \left(\frac{M}{M_\odot}\right)^{-1} \left(\frac{\xi R_\text{ch}(M)}{R_\oplus} \right) \right]^{-1},
\label{Semi-empirical mass-radius relation for hot white dwarfs}
\end{equation}
which provides a simple analytical mass-radius relation of white dwarfs, with $T_0=$ 588,862 K. The knowledge about the parameters $\mu$ and $\xi$ are the only physical ingredients to be added from the observations. 

\begin{figure}
\centering
  \includegraphics{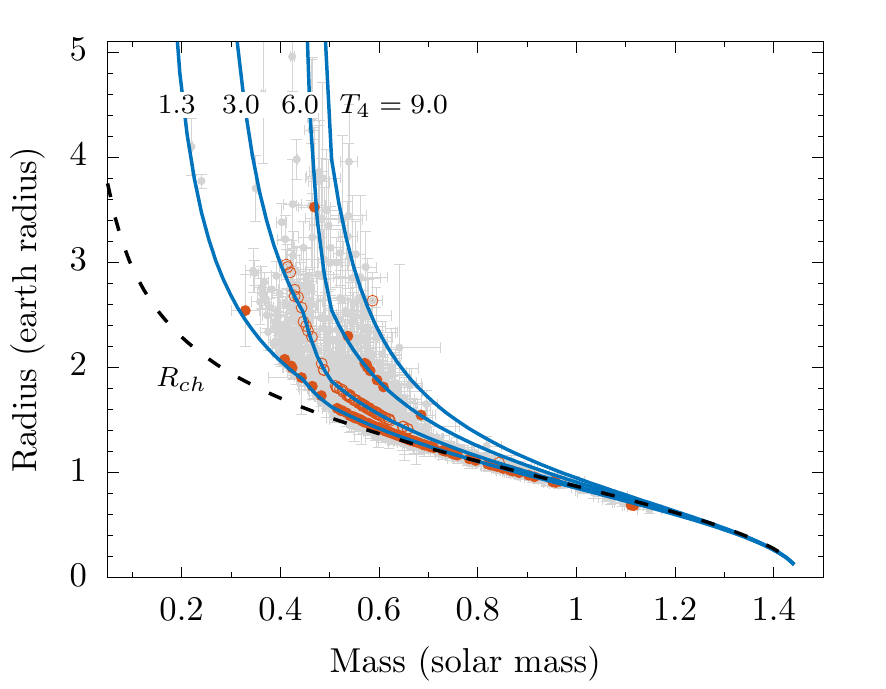}
  \caption{Semi-empirical mass-radius relation for DA-WD with different effective temperatures. The orange circles represent some values of effective temperature and the blue lines are their correspondent fits. The light gray circles are the available data for DA-WD from the SDSS-DR7. The dashed line corresponds to the Chandrasekhar model for ideal white dwarfs stars.}
  \label{fig: Relacao massa raio DA}
\end{figure}

\begin{figure}
\centering
  \includegraphics{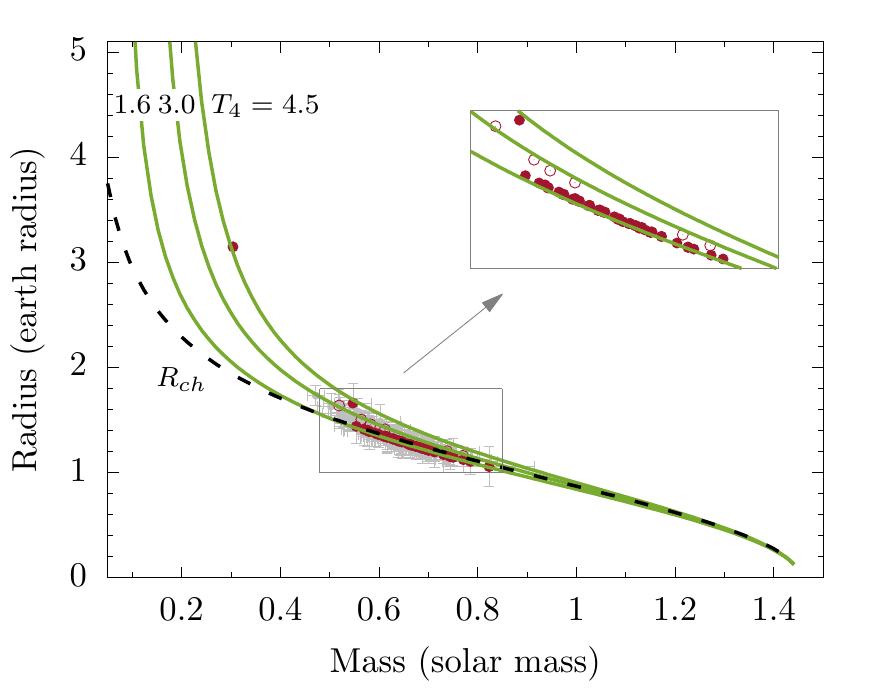}
  \caption{Semi-empirical mass-radius relation for DB-WD with different effective temperatures. The red circles represent some values of effective temperature and the green lines are their correspondent fits. The light gray circles are the available data for DB-WD from the SDSS-DR7. The dashed line corresponds to the Chandrasekhar model for ideal white dwarfs stars.}
  \label{fig: Relacao massa raio DB}
\end{figure}

For DA-WD we can use the parameters $\mu$ and $\xi$ found in Section \ref{DA white dwarfs}. The Figure \ref{fig: Relacao massa raio DA} represents this mass-radius relation of DA-WD with different effective temperatures. The data are represented in narrow ranges of effective temperature with width of $1000$ K, symbolized as the orange circles. The blue lines are the correspondent mass-radius relation obtained from Eq.\eqref{Semi-empirical radius-effective temperature relation for hot white dwarfs}, where $T_4 = T/(10^4\; \text{K})$. The Chandrasekhar mass-radius relation for CO ideal white dwarf is represented for comparison. We can note an ideal white dwarf behavior to white dwarf with mass above $1\; M_\odot$. This behavior comes from the denominator in Eq.\eqref{Semi-empirical mass-radius relation for hot white dwarfs} when the temperature parcel becomes smaller than the Chandrasekhar radius parcel. A mass-radius relation for DB white dwarfs can be obtained similarly, getting the parameters of Section \ref{DB white dwarfs}, and the result is presented in Figure \ref{fig: Relacao massa raio DB}.

A recent paper from \cite{Tremblay2016TheDwarfs} reports a sample of white dwarf parallaxes, including 4 directly observed DA-WD and other wide binaries WD. This data set can be combined with spectroscopic atmospheric parameters, as effective temperature and superficial gravity, to study the mass-radius relationship. 

\begin{figure}
\centering
  \includegraphics{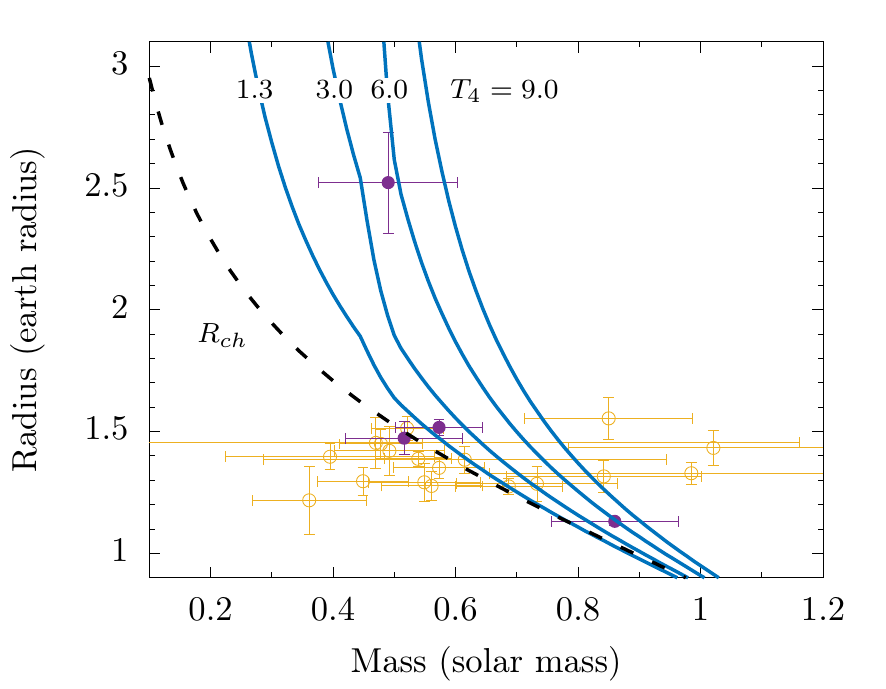}
  \caption{Semi-empirical mass-radius relation for DA-WD with different effective temperatures (blue lines). The purple solid circles correspond to the directly observed DA-WD and the yellow open circles are the DA-WD observed in wide binaries from the \emph{Gaia} DR1. The dashed line is the Chandrasekhar mass-radius relation for ideal WD.}
  \label{fig: Gaia mass-radius relation DA}
\end{figure}

Using the data from that paper, we can reproduce the estimated masses and radii  from the \emph{Gaia}-DR1 and compare with our semi-empirical mass-radius relation, as illustrated in Fig.\ref{fig: Gaia mass-radius relation DA}. The purple solid circles are the directly observed DA-WD, also identified in Table \ref{tbl-estimated masses}, and the yellow open circles are the wide binaries DA-WD. Our semi-empirical mass-radius relation is represented as the blue lines for different effective temperatures. Since our mass-radius relation is obtained from the \emph{SDSS}-DR7 and agreed very well with the directly observed DA-WD from \emph{Gaia}, we suggest that the wide binaries data be reviewed in future analysis. 

There is a observed sample of eclipsing white-dwarfs where the derivation of both mass and radius is independent. In Table \ref{tbl-estimated radii}, we calculate the radius $R$ (the last column) for the given $T_\text{eff}$ and $M=M_\text{eclipse}$ which is to be compared to the observed radius $R_\text{eclipse}$ (the fourth column) . We find that our semi-empirical mass-radius relation is in good agreement with observations.

\begin{table*}
\caption{The estimated radii for the observed DA white dwarf stars from Eclipsing Binaries.\label{tbl-estimated radii}}
\centering
\begin{minipage}{120mm}
\begin{tabular}{ccccc} 
\hline \hline
Name & $T_\text{eff}$ (K) & $M_\text{eclipse}$ ($M_\odot$) & $R_\text{eclipse}$ ($R_\oplus$)& $R$ ($R_\oplus$) \\ \hline
 CSS 41177A & 22,500(60) & 0.378(0.023) & 2.425(0.045)& 2.64(0.21) \\
 NN Ser  & 63,000(3000) & 0.535(0.012) & 2.27(0.02) & 2.31(0.12) \\
 SDSS 0857+0342 & 37,400(400) & 0.514(0.049) & 2.69(0.09) & 2.11(0.48) \\
 SDSS J1212-0123 & 17,710(40)  & 0.439(0.050) & 1.83(0.01)& 2.07(0.28) \\
 GK Vir & 50,000(670) & 0.562(0.014) & 1.85(0.03) & 1.91(0.07) \\
 QS Vir & 14,220(350) & 0.781(0.013) & 1.165(0.008)& 1.14(0.02) \\
 V471 Tau & 34,500(1000) & 0.840(0.050) & 1.17(0.08) & 1.10(0.08)\\
\hline
\end{tabular}
\end{minipage}
\end{table*}

\subsection{Effective Temperature Limit}

The star radius $R$, from Eq.\eqref{Semi-empirical mass-radius relation for hot white dwarfs}, becomes infinite when the effective temperature is equal to the limiting value, $T_\text{lim}$, given by
\begin{equation}
\frac{T_\text{lim}(M)}{T_0}=\mu \left(\frac{M}{M_\odot}\right) \left(\frac{\xi R_\text{ch}(M)}{R_\oplus} \right)^{-1}.
\label{limite temperature}
\end{equation}

Using this definition for the limiting temperature, the radius of the hot white dwarf is written as $R\propto(T_\text{lim}-T)^{-1}$.

\begin{figure}
\centering
  \includegraphics{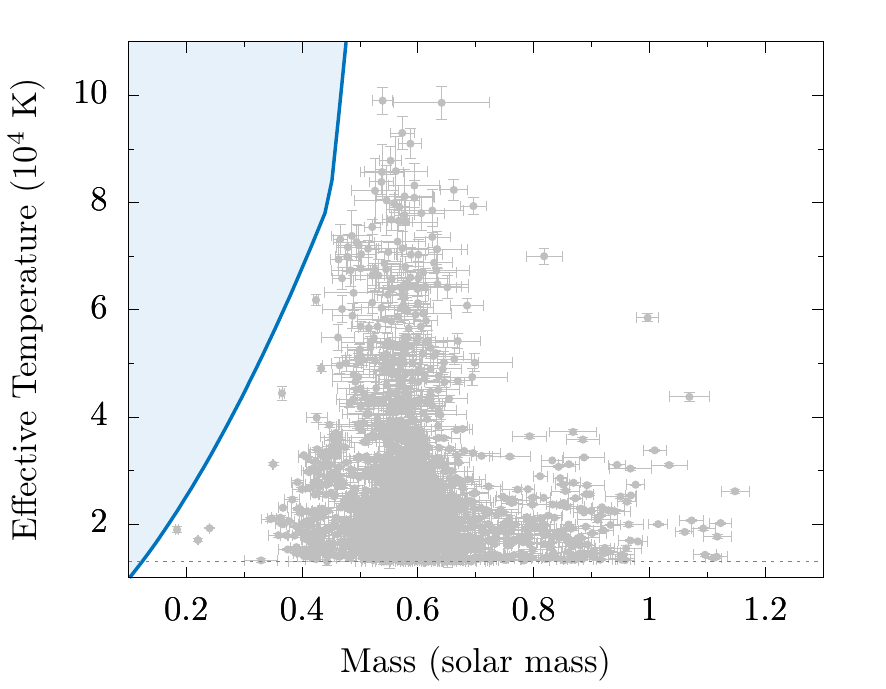}
  \caption{Effective temperature limit as a function of mass for DA-WD (blue line). The blue region represents the forbidden region of the effective temperature to the correspondent mass of DA-WD. The light gray circles are the available data for DA-WD from the SDSS-DR7. The dot line corresponds to the temperature threshold for ideal white dwarfs stars.}
  \label{fig:Temperatura Limite DA}
\end{figure}

In Fig.\ref{fig:Temperatura Limite DA}, we show $T_\text{lim}$ as function of mass $M$. The light gray circles are the effective temperature distribution as a function of mass. The blue region represents the forbidden region for the effective temperature of hot DA-WD. In fact, there is not a single point in the bulk of the region, which indicates that this temperature limit exhibits a physical behavior of data, although be a mathematical limit of our model.

\section{Central Temperature and Nuclear Ignition}
\label{sec: Central Temperature and Nuclear Ignition}

Although astronomical observables are stellar atmospheric quantities, the stellar interior quantities are of great importance to astrophysics. For instance, the central temperature and the central density determine the chemical evolution of the star and its nuclear energy generation. 

The relation between effective temperature and central temperature is given by \cite{Koester1976ConvectiveDwarfs} in the approximated form
\begin{equation}
\frac{T_\text{eff}^4}{g} = 2.05\times 10^{-10} T_c^{2.56}
\label{central temperature relation}
\end{equation}
where $g$ is the superficial gravity. This relation was obtained by fitting the data from simulations and give us a good estimate of the central temperature inside WD stars.

\begin{figure}
\centering
  \includegraphics{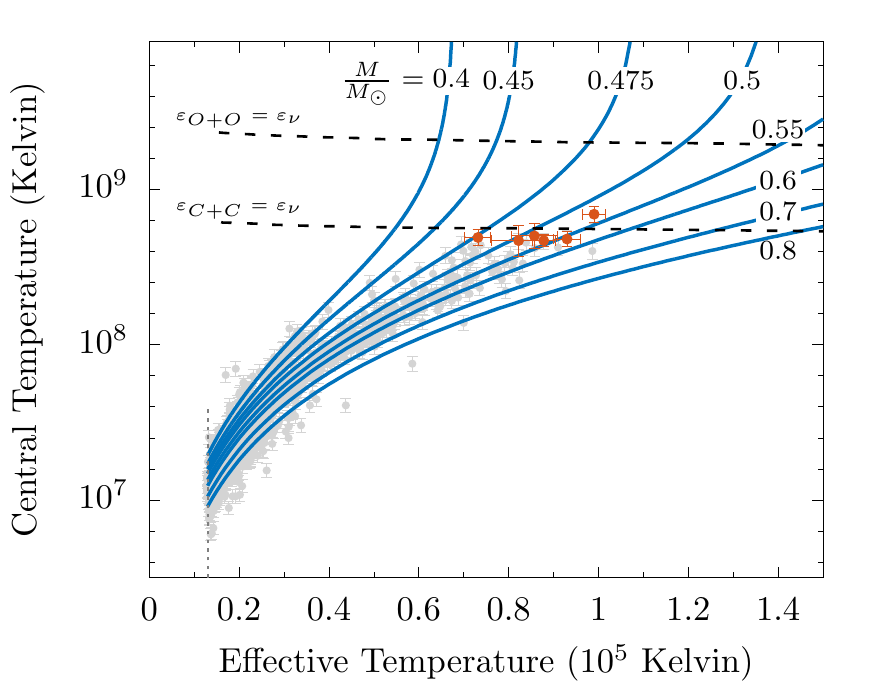}
  \caption{Central temperature as a function of the effective temperature for DA-WD with different masses (blue lines). The gray circles represent the central temperature to each DA-WD star in the available data from the SDSS-DR7, and the orange circles are the highest central temperature for WD from these data. The dashed lines are the ignition curves for carbon and oxygen fusions. The dotted line is our effective temperature threshold for hot white dwarfs.}
  \label{fig: Temperatura Central DA}
\end{figure}

We can estimate the central temperature of WD-data from SDSS-DR7, using their radii, mass and effective temperature. In Fig.\ref{fig: Temperatura Central DA} we represent these data as the gray circles.

Using our semi-empirical mass-radius relation, we can obtain the central temperature of WD using just their effective temperatures and their masses, since their radii can be calculated with this relation. The central temperature of DA-WDs as a function of their effective temperature is plotted for different mass values in Fig.\ref{fig: Temperatura Central DA}, as blue lines. 

The ignition of the nuclear material inside the white dwarf is determined by the balance between nuclear energy generation rate and local heat losses. We consider the case where the heat losses are mainly caused by neutrino emission, which is appropriate in white dwarfs, e.g., for modeling Type Ia supernova events \citep{Hillebrandt2000TypeModels}. The ignition temperature for the material depends on the central density of the star. We use the fitting formula for ignition temperature for carbon and oxygen fusions as a function of mass density, the Eq.(A.1) in \cite{Potekhin2012ThermonuclearStars}. The central densities for white dwarfs with different masses are calculated using the hydrostatic equilibrium equation with the EoS of degenerate electrons, i.e., hydrostatic equilibrium for ideal white dwarfs. This assumption is enough \emph{a posteriori} because the temperature corrections to the EoS are important just above the ignition temperature for oxygen. In Figure \ref{fig: Temperatura Central DA} we represent both carbon and oxygen ignition lines as the dashed lines, using the Eq.\eqref{central temperature relation} to estimate the correspondent effective temperature.

In Table \ref{tbl-central temperature} we present six DA-WD with central temperature (the last column) above 99\% of the carbon ignition temperature, represented by the orange circles in Fig.\ref{fig: Temperatura Central DA}. The star 200646.50-124410.9 is a special case because it is the only one WD above the carbon ignition line. The fact that this WD did not become a supernova can be understood by its internal composition, i.e., if there is more oxygen then carbon in its core, we expect no ignition inside this WD. We suggest that more careful observations and more detailed simulations must be directed to the modeling of this WD star.

\begin{table*}
\caption{The highest estimated central temperature determined for the observed DA white dwarf stars from \emph{SDSS}-DR7.\label{tbl-central temperature}}
\begin{minipage}{100mm}
\begin{tabular}{ccccc} 
\hline \hline
Name & $M$ ($M_\odot$) & $R$ ($R_\oplus$) & $T_\text{eff}$ (K) & $\log T_c$ (K) \\ \hline
 200646.50-124410.9 & 0.539(0.017) & 3.959(0.546) & 99,018(2529) & 8.84(0.05) \\
 091442.70+041455.9 & 0.538(0.037) & 3.445(0.692) & 85,714(5102) & 8.70(0.08)\\
 113303.70+290223.0 & 0.466(0.012) & 4.375(0.577) & 73,149(2867) & 8.69(0.05) \\
 102624.05+091554.8 & 0.573(0.021) & 2.957(0.335) & 92,989(3088) & 8.68(0.05) \\
 224653.73-094834.5 & 0.553(0.019) & 3.077(0.313) & 87,805(2600) & 8.67(0.04) \\
 080403.06+083030.8 & 0.526(0.041) & 3.406(0.799) & 82,219(6036) & 8.67(0.10) \\
\hline
\end{tabular}
\end{minipage}
\end{table*}

\section{Estimating Masses}
\label{sec: Estimating Masses}

One of the great achievements in white dwarf research has been the capacity to measure the effective temperatures and superficial gravities. In particular, the spectroscopic technique developed by \cite{Bergeron1992ADwarfs} for analyzing the Balmer line of hydrogen in (DA) white dwarfs has become the standard method for measuring the effective temperature and surface gravity of these stars which represent 80\% of the white dwarf population. In addition to being infrequent than their hydrogen-line DA counterparts, the hotter DB stars are characterized by an optical spectrum where the neutral helium transitions exhibit little sensitivity to effective temperature, as discussed in \cite{Bergeron2011ADWARFS}. The mass-radius relation is fundamental to compute white dwarf masses from these accurate measurements. 

Using our semi-empirical mass-radius relation, Eq.\eqref{Semi-empirical mass-radius relation for hot white dwarfs}, we can obtain the superficial gravity $g$ as a function of the mass $M$ and the effective temperature $T_\text{eff}$ of the white dwarf star according 
\begin{equation}
g(M,T_\text{eff})=\frac{GM}{R(M,T_\text{eff})^2},
\label{surface gravity}
\end{equation}
which we can be numerically invert to obtain the mass of the WD as a function of $T_\text{eff}$ and $\log{g}$.

As a test case, the most recent measurements of Sirius B from \cite{Holberg1998SiriusView} can be used to determine its mass. Sirius B is a hydrogen-rich DA-WD whose the effective temperature is 24,790(100) K and the surface gravity is $\log{g} = 8.57(0.06)$. Then, using the Eq.\eqref{surface gravity}, we can obtain the mass value of $M=0.960(0.035)\; M_\odot$. This result is close to the refined estimates of the mass $ M = 1.034(0.026)\; M_\odot$ using other method of measurement, as the \emph{Hipparcos} parallax method. 

Another example would be the PG0948+534 reported by \cite{Preval2016UnderstandingPG0948+534} as currently one of the hottest DA white dwarf stars. The authors were able to measure the $T_\text{eff}$ and the $\log{g}$ for this WD, finding $T_\text{eff} =$ 110,000 K and $\log{g} = 7.58$. For the case of PG0948+534, we find the mass value of $M = 0.640\; M_\odot$, that corroborates the hypothesis of DA hot white dwarfs with mass between $0.5-0.7\; M_\odot$ are the hottest observed DA stars, as can be seen in Figure \ref{fig: Temperatura Central DA}.

The best test is to compare our mass estimates with direct mass measurements by independent methods, such as those presented in \emph{Gaia}-DR1 by \cite{Tremblay2016TheDwarfs}. As discussed in the Section \ref{sec: Semi-Empirical Mass-Radius Relation}, the data of directly observed DA-WD are in better agreement with our model than the data of wide binaries WD. In Table \ref{tbl-estimated masses}, we estimate the mass (the last column) using the atmospheric measurements of effective temperature and superficial gravity for these WD, by Eq.\eqref{surface gravity}, and compare with the observed mass $M_\text{Gaia}$ (the fourth column).

\begin{table}
\caption{The estimated masses for the directly observed DA white dwarf stars from the \emph{Gaia}-DR1.\label{tbl-estimated masses}}
\scriptsize
\centering
\begin{tabular}{ccccc} 
\hline \hline
Name & $T_\text{eff}$ (K) & $\log g$ (cm/s$^2$) & $M_\text{Gaia}$ ($M_\odot$) & $M$ ($M_\odot$) \\ \hline
 0232+035 & 66,950(1440) & 7.40(0.07) & 0.490(0.113) & 0.518(0.013) \\
 1314+293 & 56,800(1250) & 7.89(0.07) & 0.516(0.096) & 0.644(0.028) \\
 1647+591 & 12,510(200) & 8.34(0.05) & 0.860(0.103) & 0.807(0.031) \\ 
 2117+539 & 14,680(240) & 7.91(0.05) & 0.573(0.071) & 0.561(0.025) \\ 
\hline
\end{tabular}
\end{table}

\section{Discussion and Conclusions}
\label{sec: Conclusions}

In this paper, introducing a very simple model for the outer layer of hot WDs we analyzed the \emph{SDSS}-DR7 and derived a simple, analytic semi-phenomenological relation among effective temperature, mass and radius of hot white dwarfs, the Eq.\eqref{Semi-empirical radius-effective temperature relation for hot white dwarfs}.

From this relation, we observe that there are two essential differences between hydrogen-rich DA white dwarf and hydrogen-deficient DB white dwarf: their outer layer composition and their effective temperature range. 

As discussed in \cite{HotWD2011}, the DA-WD are much easier to classify because the Balmer lines of hydrogen across a wide range of effective temperature $T_\text{eff}$, from 4,000 up to 120,000 K and higher, whereas DB-WD exhibit He I lines but with a lower effective temperature range, from 12,000 to 45,000 K. Confirming the different effective temperature ranges for DA and DB. 

The difference in the outer layer composition was presented by the parameter $\mu$ of both DA and DB, and it indicates that the parameter is closely related to the mean molecular weight of this region, whose the information give us clues about the chemical composition of the material.  

The parameter $\xi$ must be related to the Rosseland optical-depth mentioned in \cite{Baschek1991} and it depends on the chemical composition of the material, due to the opacity of the material. The fact of $\xi\lesssim 1$ shows that the region responsible for the photon emission is essentially in the border of the core described by the degenerate electron gas, suggesting that a small portion of this surface is melted into the outer layer. 

Our result permits us to obtain a mass-radius relation, the Eq.\eqref{Semi-empirical mass-radius relation for hot white dwarfs}, and estimates of radii for WDs for known mass and temperature with other methods. Furthermore, our formula exhibits a mathematical limit to the effective temperature, and curiously there is not a single white dwarf star in the bulk of the forbidden region imposed by this limit. A further study to understand the existence of such a limiting temperature is required.

The central temperature can be evaluated using the relation between effective temperature and superficial gravity derived from \cite{Koester1976ConvectiveDwarfs} using numerical models of WD. The data from \emph{SDSS}-DR7 present six DA-WD with central temperatures very close to the carbon ignition temperature. There is only one DA-WD with central temperature above the ignition temperature. If our analytic expression reflects the physical systematics correctly, we may think of the possibility that the core of this WD is composed by oxygen instead of carbon. Numerical simulations and future observations are required for the better understanding whether such WD is a possible Type Ia supernova progenitor.

The mass-radius relation obtained in this work allows us to obtain mass estimates from atmospheric measurements of effective temperature and superficial gravity. We use this method to estimate masses of the well known Sirius B and other DA-WD from the recent \emph{Gaia}-DR1. Although they are distinct methods, our mass evaluations are in good accordance with the masses measured by \emph{Gaia}, considering their uncertainties. This result confirm our relation, Eq.\eqref{Semi-empirical radius-effective temperature relation for hot white dwarfs}, as a great constraining for effective temperature, mass and radius of hot white dwarfs.

\section*{Acknowledgements}

The author acknowledges the members of ICE group of the Institute of Physics for fruitful discussions and comments. In particular, the author would like to thank Profs. T. Kodama and J. R. T. de Mello Neto for reading the manuscript and their suggestions. This work is financially supported by CNPq.




\bibliographystyle{mnras}
\bibliography{Mendeley} 

\begin{thebibliography}{}
\makeatletter
\relax
\def\mn@urlcharsother{\let\do\@makeother \do\$\do\&\do\#\do\^\do\_\do\%\do\~}
\def\mn@doi{\begingroup\mn@urlcharsother \@ifnextchar [ {\mn@doi@}
  {\mn@doi@[]}}
\def\mn@doi@[#1]#2{\def\@tempa{#1}\ifx\@tempa\@empty \href
  {http://dx.doi.org/#2} {doi:#2}\else \href {http://dx.doi.org/#2} {#1}\fi
  \endgroup}
\def\mn@eprint#1#2{\mn@eprint@#1:#2::\@nil}
\def\mn@eprint@arXiv#1{\href {http://arxiv.org/abs/#1} {{\tt arXiv:#1}}}
\def\mn@eprint@dblp#1{\href {http://dblp.uni-trier.de/rec/bibtex/#1.xml}
  {dblp:#1}}
\def\mn@eprint@#1:#2:#3:#4\@nil{\def\@tempa {#1}\def\@tempb {#2}\def\@tempc
  {#3}\ifx \@tempc \@empty \let \@tempc \@tempb \let \@tempb \@tempa \fi \ifx
  \@tempb \@empty \def\@tempb {arXiv}\fi \@ifundefined
  {mn@eprint@\@tempb}{\@tempb:\@tempc}{\expandafter \expandafter \csname
  mn@eprint@\@tempb\endcsname \expandafter{\@tempc}}}

\bibitem[\protect\citeauthoryear{Althaus, Garc{\'{i}}a-Berro, Isern  \&
  C{\'{o}}rsico}{Althaus et~al.}{2005}]{Althaus2005Mass-radiusStars}
Althaus L.~G.,  Garc{\'{i}}a-Berro E.,  Isern J.,   C{\'{o}}rsico A.~H.,  2005,
  \mn@doi [Astronomy and Astrophysics] {10.1051/0004-6361:20052996}, 441, 689

\bibitem[\protect\citeauthoryear{Althaus, Panei, Romero, Rohrmann,
  C{\'{o}}rsico, Garc{\'{i}}a-Berro  \& Miller~Bertolami}{Althaus
  et~al.}{2009a}]{Althaus2009EvolutionProgenitors}
Althaus L.~G.,  Panei J.~a.,  Romero a.~D.,  Rohrmann R.~D.,  C{\'{o}}rsico
  a.~H.,  Garc{\'{i}}a-Berro E.,   Miller~Bertolami M.~M.,  2009a, \mn@doi
  [Astronomy and Astrophysics] {10.1051/0004-6361/200911640}, 502, 207

\bibitem[\protect\citeauthoryear{Althaus, Panei, Miller~Bertolami,
  Garc{\'{i}}a-Berro, C{\'{o}}rsico, Romero, Kepler  \& Rohrmann}{Althaus
  et~al.}{2009b}]{Althaus2009NEWEVOLUTION}
Althaus L.~G.,  Panei J.~A.,  Miller~Bertolami M.~M.,  Garc{\'{i}}a-Berro E.,
  C{\'{o}}rsico A.~H.,  Romero A.~D.,  Kepler S.~O.,   Rohrmann R.~D.,  2009b,
  \mn@doi [The Astrophysical Journal] {10.1088/0004-637X/704/2/1605}, 704, 1605

\bibitem[\protect\citeauthoryear{Althaus, C{\'{o}}rsico, Isern  \&
  Garc{\'{i}}a-Berro}{Althaus et~al.}{2010}]{Althaus2010EvolutionaryStars}
Althaus L.~G.,  C{\'{o}}rsico A.~H.,  Isern J.,   Garc{\'{i}}a-Berro E.,  2010,
  \mn@doi [The Astronomy and Astrophysics Review] {10.1007/s00159-010-0033-1},
  18, 471

\bibitem[\protect\citeauthoryear{Baschek, Scholz  \& Wehrse}{Baschek
  et~al.}{1991}]{Baschek1991}
Baschek B.,  Scholz M.,   Wehrse R.,  1991, Astronomy and Astrophysics, 246,
  374

\bibitem[\protect\citeauthoryear{Bergeron, Saffer  \& Liebert}{Bergeron
  et~al.}{1992}]{Bergeron1992ADwarfs}
Bergeron P.,  Saffer R.~A.,   Liebert J.,  1992, \mn@doi [The Astrophysical
  Journal] {10.1086/171575}, 394, 228

\bibitem[\protect\citeauthoryear{Bergeron et~al.,}{Bergeron
  et~al.}{2011}]{Bergeron2011ADWARFS}
Bergeron P.,  et~al., 2011, \mn@doi [The Astrophysical Journal]
  {10.1088/0004-637X/737/1/28}, 737, 28

\bibitem[\protect\citeauthoryear{Boshkayev, Rueda, Zhami, Kalymova  \&
  Balgymbekov}{Boshkayev et~al.}{2016}]{Boshkayev2016EquilibriumTemperatures}
Boshkayev K.~A.,  Rueda J.~a.,  Zhami B.~A.,  Kalymova Z.~A.,   Balgymbekov
  G.~S.,  2016, \mn@doi [International Journal of Modern Physics: Conference
  Series] {10.1142/S2010194516601290}, 41, 1660129

\bibitem[\protect\citeauthoryear{Chandrasekhar}{Chandrasekhar}{1931}]{Chandrasekhar1931TheMass}
Chandrasekhar S.,  1931, Monthly Notices of the Royal Astronomical Society, 91,
  456

\bibitem[\protect\citeauthoryear{Eisenstein et~al.,}{Eisenstein
  et~al.}{2006}]{Eisenstein2006A4}
Eisenstein D.~J.,  et~al., 2006, \mn@doi [The Astrophysical Journal Supplement
  Series] {10.1086/507110}, 167, 40

\bibitem[\protect\citeauthoryear{Hamada \& Salpeter}{Hamada \&
  Salpeter}{1961}]{Hamada1961ModelsStars.}
Hamada T.,  Salpeter E.~E.,  1961, \mn@doi [The Astrophysical Journal]
  {10.1086/147195}, 134, 683

\bibitem[\protect\citeauthoryear{Hillebrandt \& Niemeyer}{Hillebrandt \&
  Niemeyer}{2000}]{Hillebrandt2000TypeModels}
Hillebrandt W.,  Niemeyer J.~C.,  2000, \mn@doi [Annual Review of Astronomy and
  Astrophysics] {10.1146/annurev.astro.38.1.191}, 38, 191

\bibitem[\protect\citeauthoryear{Hillebrandt, Kromer, R{\"{o}}pke  \&
  Ruiter}{Hillebrandt et~al.}{2013}]{Hillebrandt2013}
Hillebrandt W.,  Kromer M.,  R{\"{o}}pke F.~K.,   Ruiter A.~J.,  2013, \mn@doi
  [Frontiers of Physics] {10.1007/s11467-013-0303-2}, 8, 116

\bibitem[\protect\citeauthoryear{Holberg, Barstow, Bruhweiler, Cruise  \&
  Penny}{Holberg et~al.}{1998}]{Holberg1998SiriusView}
Holberg J.~B.,  Barstow M.~A.,  Bruhweiler F.~C.,  Cruise A.~M.,   Penny A.~J.,
   1998, \mn@doi [The Astrophysical Journal] {10.1086/305489}, 497, 935

\bibitem[\protect\citeauthoryear{Holberg, Oswalt  \& Barstow}{Holberg
  et~al.}{2012}]{Holberg2012OBSERVATIONALRELATION}
Holberg J.~B.,  Oswalt T.~D.,   Barstow M.~A.,  2012, \mn@doi [The Astronomical
  Journal] {10.1088/0004-6256/143/3/68}, 143, 68

\bibitem[\protect\citeauthoryear{Hubbard \& Wagner}{Hubbard \&
  Wagner}{1970}]{Hubbard1970HotDwarfs}
Hubbard W.~B.,  Wagner R.~L.,  1970, \mn@doi [The Astrophysical Journal]
  {10.1086/150292}, 159, 93

\bibitem[\protect\citeauthoryear{Kepler, Kleinman, Nitta, Koester, Castanheira,
  Giovannini, Costa  \& Althaus}{Kepler et~al.}{2007}]{Kepler2007}
Kepler S.~O.,  Kleinman S.~J.,  Nitta A.,  Koester D.,  Castanheira B.~G.,
  Giovannini O.,  Costa a. F.~M.,   Althaus L.~G.,  2007, \mn@doi [Monthly
  Notices of the Royal Astronomical Society]
  {10.1111/j.1365-2966.2006.11388.x}, 375, 1315

\bibitem[\protect\citeauthoryear{Kippenhahn, Weigert  \& Weiss}{Kippenhahn
  et~al.}{2012}]{Kippenhahn2012}
Kippenhahn R.,  Weigert A.,   Weiss A.,  2012, {Stellar Structure and
  Evolution}, 2 edn.
Astronomy and Astrophysics Library, Springer-Verlag Berlin Heidelberg

\bibitem[\protect\citeauthoryear{Kleinman et~al.,}{Kleinman
  et~al.}{2013}]{Kleinman2013}
Kleinman S.~J.,  et~al., 2013, \mn@doi [The Astrophysical Journal Supplement
  Series] {10.1088/0067-0049/204/1/5}, 204, 5

\bibitem[\protect\citeauthoryear{Koester}{Koester}{1976}]{Koester1976ConvectiveDwarfs}
Koester D.,  1976, Astronomy and Astrophysics, 52, 415

\bibitem[\protect\citeauthoryear{Koester \& Chanmugam}{Koester \&
  Chanmugam}{1990}]{Koester1990}
Koester D.,  Chanmugam G.,  1990, \mn@doi [Reports on Progress in Physics]
  {10.1088/0034-4885/53/7/001}, 53, 837

\bibitem[\protect\citeauthoryear{Marshak}{Marshak}{1940}]{Marshak1940TheStars.}
Marshak R.~E.,  1940, \mn@doi [The Astrophysical Journal] {10.1086/144225}, 92,
  321

\bibitem[\protect\citeauthoryear{Nauenberg}{Nauenberg}{1972}]{Nauenberg1972AnalyticStars}
Nauenberg M.,  1972, \mn@doi [The Astrophysical Journal] {10.1086/151568}, 175,
  417

\bibitem[\protect\citeauthoryear{Potekhin \& Chabrier}{Potekhin \&
  Chabrier}{2012}]{Potekhin2012ThermonuclearStars}
Potekhin A.~Y.,  Chabrier G.,  2012, \mn@doi [Astronomy {\&} Astrophysics]
  {10.1051/0004-6361/201117938}, 538, A115

\bibitem[\protect\citeauthoryear{Preval \& Barstow}{Preval \&
  Barstow}{2016}]{Preval2016UnderstandingPG0948+534}
Preval S.~P.,  Barstow M.~A.,  2016, e-print arXiv:1610.01677, pp~1--6

\bibitem[\protect\citeauthoryear{Shapiro \& Teukolsky}{Shapiro \&
  Teukolsky}{1983}]{Shapiro1983}
Shapiro S.~L.,  Teukolsky S.~A.,  1983, {Black Holes, White Dwarfs and Neutron
  Stars: The Physics of Compact Objects}, 1 edn.
Wiley-VCH

\bibitem[\protect\citeauthoryear{Sion}{Sion}{2011}]{HotWD2011}
Sion E.~M.,  2011, in Hoard D. D.~W.,  ed., , White Dwarf Atmospheres and
  Circumstellar Environments.
Wiley-VCH Verlag GmbH {\&} Co. KGaA, Weinheim, Germany, Chapt.~1, pp 1--23,
  \mn@doi{10.1002/9783527636570.ch1}, \url
  {http://dx.doi.org/10.1002/9783527636570.ch1}

\bibitem[\protect\citeauthoryear{Tremblay et~al.,}{Tremblay
  et~al.}{2016}]{Tremblay2016TheDwarfs}
Tremblay P.~E.,  et~al., 2016, e-print arXiv:1611.00629, pp 1--13

\bibitem[\protect\citeauthoryear{Weiss, Hillebrandt, Thomas  \& Ritter}{Weiss
  et~al.}{2004}]{Cox2004}
Weiss A.,  Hillebrandt W.,  Thomas H.-C.,   Ritter H.,  2004, {Cox and Giuli's
  Principles of Stellar Structure}, 2 edn.
Advances in Astronomy {\&} Astrophysics, Cambridge Scientific Publishers

\bibitem[\protect\citeauthoryear{de Carvalho, Rotondo, Rueda  \&
  Ruffini}{de~Carvalho et~al.}{2014}]{DeCarvalho2014}
de Carvalho S.~M.,  Rotondo M.,  Rueda J.~A.,   Ruffini R.,  2014, \mn@doi
  [Physical Review C] {10.1103/PhysRevC.89.015801}, 89, 015801

\makeatother
\end{thebibliography}



\bsp	
\label{lastpage}
\end{document}